\documentclass[preprint,pra,aps,draft,showpacs]{revtex4}
\usepackage{graphicx}% Include figure files
\usepackage{dcolumn}% Align table columns on decimal point
\usepackage{bm}% bold math

\def\E{{\cal E}}
\def\x{{\bf x}}
\def\y{\mathbf{y}}
\def\k{\mathbf{k}}
\def\q{\mathbf{q}}

\def\P{\mathbf{P}}
\def\r{\mathbf{r}}
\def\a{\alpha}
\def\b{\beta}
\def\om{\omega}

\def\ih{{ \frac{i}{\hbar} }}
\def\au{{\underline \alpha}}
\def\bu{{\underline \beta}}
\def\half{\frac{1}{2}}

\begin{document}

%\preprint{APS/123-QED}

\title{Decoherence and Records for the Case of \\ a Scattering Environment}

\author{P.J.Dodd}
\author{J.J.Halliwell}%
\affiliation{Blackett Laboratory \\ Imperial College \\ London SW7
2BZ \\ UK }

%\affiliation{


\begin{abstract}
Using non-relativistic many body quantum field theory, a master
equation is derived for the reduced density matrix of a dilute gas
of massive particles undergoing scattering interactions with an
environment of light particles. The dynamical variable that
naturally decoheres (the pointer basis) is essentially the local
number density of the dilute gas. Earlier master equations for
this sort of system (such as that derived by Joos and Zeh) are
recovered on restricting to the one-particle sector for the
distinguished system. The derivation shows explicitly that the
scattering environment stores information about the system by
``measuring'' the number density. This therefore provides an
important example of the general connection between decoherence
and records indicated by the decoherent histories approach to
quantum theory. It also brings the master equation for this system
into a form emphasizing the role of local densities, which is
relevant to current work on deriving hydrodynamic equations from
quantum theory.
\end{abstract}

\pacs{03.65.-w, 03.65.Yz, 03.65.Ta, 05.70.Ln}% PACS, the Physics and Astronomy
\maketitle

\section{Introduction}

The notion of decoherence plays a central role in studies of
emergent classicality and the foundations of quantum theory
generally \cite{Zur1}. While it is usually regarded as signifying
the loss of quantum coherence for a system of interest, it may
also usefully be regarded as an indication of the degree to which
information about the system is stored somewhere in the system or
in its immediate environment \cite{GeH1,Hal1}. It is in this way
that decoherence is related to ``generalized measurements''. An
important application of these ideas is in quantum cosmology
\cite{Har1}. There, in applying quantum theory to the very early
universe, there are no actual measuring device to measure what was
happening. The process of decoherence, however, guarantees that
measurements we make in the present are correlated with
alternatives in the past.

These ideas are perhaps most transparent when formulated in terms
of the decoherent histories approach to quantum theory
\cite{GeH1,GeH2,Gri,Omn,Har1,Hal2}.  Other approaches to
decoherence, such as Zurek's ``einselection'' approach
\cite{Zur1,Zur2,Zur4}, related density matrix approaches
\cite{JoZ} or quantum state diffusion \cite{GiP,HaZ}, may be
equally useful for analyzing these issues, but will not be
explored here. It is the aim of this paper, continuing in the
spirit of Ref.\cite{Hal1}, to investigate the connection between
decoherence and information storage. To fix ideas, we briefly
review the decoherent histories approach (although the general
results of this paper are by no means specific that approach).

In the decoherent histories approach
\cite{GeH1,GeH2,Gri,Omn,Har1,Hal2}, probabilities are assigned to
histories via the formula,
\begin{equation}
p (\a_1, \a_2, \cdots ) = {\rm Tr} \left( C_{\au} \rho
C_{\au}^{\dag} \right)
\label{1.1}
\end{equation}
where $C_{\au} $ denotes a time-ordered string of projectors at
times $t_1 \cdots t_n$,
\begin{equation}
C_{\au}= P_{\a_n} e^{ -\ih H (t_n - t_{n-1}) } P_{\a_{n-1}} \cdots
e^{ -\ih H (t_2 - t_1) } P_{\a_1} \label{1.2}
\end{equation}
and $ \au $ denotes the string $\a_1, \a_2, \cdots \a_n $. We are
interested in sets of histories which satisfy the condition of
decoherence, which is that decoherence functional
\begin{equation}
D(\au,\au') = {\rm Tr} \left( C_{\au} \rho C_{\au'}^{\dag} \right)
\label{1.3}
\end{equation}
is zero when $\au \ne \au' $. Decoherence implies the weaker
condition that $ {\rm Re} D(\au,\au') = 0 $ for $\au \ne \au' $,
and this is equivalent to the requirement that the above
probabilities satisfy the probability sum rules.

The stronger condition of decoherence is the more interesting one
since it is related to the existence of records -- information
storage about the histories somewhere in the system. More
precisely, if the initial state is pure, decoherence means that
there exist a set of alternatives at the final time $t_n $ which
are perfectly correlated with the alternatives $\a_1 \cdots \a_n $
at times $t_1 \cdots t_n $ \cite{GeH2,GeH3}. This follows because,
with a pure initial state $| \Psi \rangle$, the decoherence
condition implies that the states $ C_{\au} | \Psi \rangle $ are
an orthogonal set. It is therefore possible to introduce a
projection operator $ R_{\bu} $ (which is generally not unique)
such that
\begin{equation}
R_{ \bu } C_{\au} | \Psi \rangle = \delta_{\au \bu} C_{\au} | \Psi
\rangle \label{1.4}
\end{equation}
It follows that the extended histories characterized by the chain
$ R_{\bu} C_{\au} | \Psi \rangle $ are decoherent, and one can
assign a probability to the histories $\au$ and the records $\bu$,
given by
\begin{equation}
p( \a_1, \a_2, \cdots \a_n ; \b_1, \b_2 \cdots \b_n ) = {\rm Tr}
\left( R_{\b_1 \b_2 \cdots \b_n} C_{\au} \rho C_\au^{\dag} \right)
\label{1.5}
\end{equation}
This probability is then zero unless $\a_k = \b_k $ for all $k$,
in which case it is equal to the original probability  $ p(\a_1,
\cdots \a_n) $. Hence either the $\a$'s or the $\b$'s can be
completely summed out of Eq.(\ref{1.5}) without changing the
probability, so the probability for the histories can be entirely
replaced by the probability for the records at a fixed moment of
time at the end of the history:
\begin{equation}
p (\au ) = {\rm Tr} \left( R_{\au}  \rho (t_n ) \right) = {\rm Tr}
\left( C_{\au} \rho C_{\au}^{\dag} \right)
\label{1.6}
\end{equation}
Conversely, the existence of records $\b_1, \cdots \b_n $ at some
final time perfectly correlated with earlier alternatives $\a_1,
\cdots \a_n $ at $t_1, \cdots t_n $ implies decoherence of the
histories.

These issues are most usefully investigated in the context of
particular models, and it then becomes possible to ask some more
precise questions: Which dynamical variables in the environment
store the information about the decoherent histories? Or what is
essentially the same thing, how are the ``pointer basis''
variables stored in the environment? How is the amount of
decoherence related to the amount of information stored?

Ref.\cite{Hal1} investigated these questions in the context
of the quantum Brownian model (QBM), which
consists of a particle of large mass $M$ moving in a potential
$V(x)$ and linearly coupled to a bath of harmonic oscillators. The
total system is therefore described by the action,
\begin{eqnarray}
S_T [x(t), q_n(t)] &=& \int dt \left[ \half M \dot x^2
- V(x) \right]
\nonumber \\
& + &\sum_n \int dt \left[ \half m_n \dot q_n^2
- \half m_n \om_n^2 q_n^2 - c_n q_n x \right]
\label{1.7}
\end{eqnarray}
In the traditional discussion of decoherence in this model, it is
shown that for a continuum of oscillators in a thermal state, the
influence functional or density matrix become approximately
diagonalized in position. This may be seen, for example, through
the master equation for the reduced density matrix $\rho (x,y) $ of the
distinguished system \cite{CaL}, which in the high temperature limit is
\begin{eqnarray}
{\partial \rho \over \partial t} & = &
{i \hbar \over 2 m} \left( { \partial^2 \rho \over \partial x^2}
- {\partial^2 \rho \over \partial y^2} \right)
- { i \over \hbar} \left( V(x) - V(y) \right) \rho
\nonumber \\
& - & \gamma (x - y ) \left( { \partial \rho \over \partial x}
- { \partial \rho \over \partial y } \right)
- { 2 m \gamma k T  \over \hbar^2 } (x-y)^2 \rho
\label{1.8}
\end{eqnarray}
In Ref.\cite{Hal1}, the issue of how the information about
position is stored in the environment was addressed. The system is
linear in the oscillators, so the classical and quantum dynamics
coincide for the environment. Classically, the equations of motion
of the environment of oscillators are,
\begin{equation}
m_n \ddot q_n + m_n \omega_n^2 q_n = - c_n x(t)
\label{1.9}
\end{equation}
The solution to this equation, with fixed $p_n (0)$, $q_n (0) $ is
\begin{eqnarray}
q_n (\tau ) & = & q_n (0) \cos \omega_n \tau + { p_n (0 )
\over m_n \omega_n } \sin \omega_n \tau
\nonumber \\
\quad \quad \quad & - &
{c_n \over m_n \om_n } \int_0^\tau dt  \ x(t) \ \sin \omega_n
(\tau -t)
\\
p_n (\tau ) & = & { p_n (0 ) } \cos \omega_n
\tau - m_n \omega_n q_n (0) \sin \omega_n \tau
\nonumber \\
\quad \quad \quad & - & c_n  \int_0^\tau dt \ x(t) \ \cos \omega_n (\tau -t)
\label{1.10}
\end{eqnarray}
where $p_n = m \dot q_n$. From this solution, one can see that at
the final time $\tau $, the positions and momenta of the
environment of oscillators depend on the particle's trajectory
$x(t)$ via the temporally non-local quantities
\begin{eqnarray}
X^s_n  & = & \int_0^\tau dt  \ x(t) \ \sin \omega_n (\tau
-t)
\label{1.11a} \\
X^c_n  & = & \int_0^\tau dt  \ x(t) \ \cos \omega_n
(\tau -t)
\label{1.11b}
\end{eqnarray}
These are essentially the Fourier modes of the particle's
trajectory $x(t)$. Hence, each oscillator stores a single Fourier
mode of the trajectory, and therefore by using a large number of
oscillators, information about many Fourier modes is stored from
which the approximate trajectory may be recovered. Furthermore,
since it is the Fourier modes that are naturally registered in the
environment, rather than positions at each moment of time,
decoherence is in fact most clearly seen in terms of the variables
(\ref{1.11a}), (\ref{1.11b}), rather than position, as shown in
Ref.\cite{Hal1}.
The variables are non-local in
time so it can only be seen at the level of an influence
functional expressed in path integral language, rather than a
master equation. Hence, in this model, it was possible to see
exactly how the environment stored the information about the
system's trajectory in configuration space. Furthermore, a
detailed quantitative estimate of the information content was also
carried out in Ref.\cite{Hal1}.

Although a very illustrative model, the quantum Brownian motion
model is not the most relevant model for decoherence in
physically interesting situations. Far more physically significant
is the case in which the environment is a set of light particles
which interact with the distinguished particle by a scattering
process. The resulting master equation, first derived by Joos and
Zeh, is very similar in form to the QBM case, Eq.(\ref{1.8})
\cite{JoZ,Dio,GaF}. But the dynamics of the environment, and
therefore the means of information storage, are rather different.

The aim of this paper is to investigate the connection between
decoherence and records in the case of decoherence by a scattering
environment. In some ways it is simpler, since, in the usual
assumption of widely separated timescales for system and
environment dynamics, each environmental particle scatters briefly
off the distinguished particles, and moves freely thereafter,
carrying some information about the distinguished particles. This
process can therefore be described by a Markovian master equation,
and the process of information storage and decoherence may be
described in a moment by moment manner (unlike the quantum Brownian
motion case, where
the environment oscillators store information about the entire
history). This means in fact that we do not need to make use of the
full machinery of the decoherence functional -- it is sufficient in fact
to look at the evolution of the reduced density operator.

In the quantum Brownian motion model case, the system variables
the environment measures were actually identified quite simply,
from the classical equations of motion. In the scattering case,
the system variables measured most directly by the environment are
also determined quite easily, by examining simple scattering
processes. In particular, suppose we consider the scattering
of some light particles off a dilute gas of a set of more massive
particles with coordinates $\q_j$.
Then it follows quite straightforwardly from simple scattering
theory (and we will in fact demonstrate this) that the scattering
amplitude is proportional to
the Fourier transform of the number density of the massive
particles,
\begin{equation}
N_{\k} = \sum_j e^{ i \k \cdot \q_ j }
\label{1.12}
\end{equation}
This means that, loosely speaking, for a known
interaction potential, measurements of the initial and final
momenta of the scattering environment determine the number density
of the distinguished system.

Of course, the number density is closely related to position,
which is normally held to be the preferred basis in these
calculations. But, following the lead of the oscillator model, we
expect decoherence to look simplest in terms of the dynamical
variables which are most simply and directly stored in the
environment. Our aim is therefore to give a derivation of the
master equation which emphasize the central role played by the
number density. We have found that the derivation is in fact most
transparent in terms of non-relativistic many body quantum field
theory, where the number density appears very naturally. We will
give an alternative and more general derivation of the master
equation, using many body theory, which brings out the role of
local number density more clearly, hence showing the connection
with records.

It is pertinent at this stage to mention the Lindblad form of
the master equation \cite{Lin}, which is the most general possible form a
master equation can take under the assumption that the evolution
is Markovian (a condition well-satisfied in a wide variety of
interesting models). The Lindblad master equation is
\begin{equation}
\frac { d \rho}{ dt} = -i  [H, \rho] - \half \sum_{j=1}^n \left( \{
L_j^{\dag} L_j, \rho \} - 2 L_j \rho L_j^{\dag} \right)
\label{1.13}
\end{equation}
Here, $H$ is the Hamiltonian of the distinguished subsystem
(sometimes modified by terms depending on the $L_j$) and the $n$
operators $L_j$ model the effects of the environment. The master
equation of quantum Brownian motion, for example, is of this form
with
\begin{equation}
L = \left( \frac {4 m \gamma k T } { \hbar^2} \right)^{\half} x + i \left( \frac {\gamma}
{2 m k T } \right)^{\half} p
\label{1.14}
\end{equation}
as described in Refs.\cite{Dio2,HaZ}. (Actually, the master equation (\ref{1.8})
is not strictly of the Lindblad form, and as a consequence can suffer from a violation
of positivity \cite{Amb}. However, the difference between Eq.(\ref{1.8})
and the Lindblad form with $L$ given by Eq.(\ref{1.14}) is of order $1/ T$ which
does not matter for high temperatures).

The Lindblad operators $L_j$
determine the basis in which the density operator tends to become
approximately diagonal, or what is essentially the same, the sets
of variables describing an approximately decoherent set of
histories. This may be seen from the formal solution to the
Lindblad equation \cite{DGHP}.
Consider the case of a single
Lindblad operator $L = L_R + i L_I $, where $L_R$, $L_I$ are hermitian.
Divide the finite time interval $[0,t]$
into $K$ subintervals, so that $ t = K  \delta t $, and let $\delta t \rightarrow
0$, $ K \rightarrow \infty $, holding $t$ constant. Then, the formal
solution to the Lindblad equation is obtained by taking the
limit $\delta t \rightarrow 0 $, $K \rightarrow \infty $ (with $t$ fixed)
of the expression,
\begin{eqnarray}
\rho(t ) & = &
\ \left( { \delta t \over \pi } \right)^K
\int  d^2\l_1 \cdots d^2\l_K
\nonumber \\ & \times &
\prod_{m=1}^K
\ \exp \left( {\delta t \over 2} ( \ell^*_m L - \ell_m L^{\dag} ) \right)
\ \exp \left( - {\delta t \over 2 } \ |L -\ell_m|^2 \right)
\ \exp \left( - \ih H' \delta t \right)
\rho(0) \nonumber \\
& \times &
\prod_{m=1}^K
\ \exp \left( \ih H'\delta t \right)
\ \exp \left( - { \delta t \over 2 } \ |L  -\ell_m|^2 \right)
\ \exp \left( - { \delta t \over 2 } (\ell^*_m L - \ell_m L^{\dag} )\right)
\end{eqnarray}
Here, $H' = H + {i \hbar \over 4} [L, L^{\dag}]$,
and the $\ell_m$ are complex numbers at the discrete moments of time
labelled by $m$. We use the notation
\begin{equation}
|L-\ell_m|^2\equiv
\left(L_R-{\rm Re} \ell_m\right)^2+\left(L_I-{\rm Im}\ell_m\right)^2
\end{equation}
The ordering of the
operators at each moment of time
is irrelevant in the limit $\delta t \rightarrow 0 $
(although the operators at different times are time-ordered,
according to increasing $m$). That this is the solution is
readily verified by explicit computation. The solution has the form
of a ``measurement process" of the $L$'s, continuous in time,
with ``feedback" via the terms $ (\ell^*_m L - \ell_m L^{\dag} )$
\cite{Caves}. That is, one can see that the effect of the environment is to
produce a tendency towards diagonality in $L$.

We shall show that a many-body theory derivation of the master
equation in the case of a scattering environment leads to a master
equation of the Lindblad form (under the assumption that the environment
dynamics are much faster than the system dynamics), and that the
Lindblad operators are essentially the local number density.
The previous forms of the master equation are recovered in the
one-particle sector for the system of massive particles.

This work grew out of a more ambitious programme,
in the context of the
decoherent histories approach to quantum theory, which aims to
give a very general account of emergent classicality.
In particular, it is asserted that at sufficiently coarse-grained
scales, the local densities (number, momentum, energy)
define a a set of habitually decohering variables, even in the absence
of an environment \cite{GeH1,Hal3}. This is because they are locally
conserved, and therefore slowly varying when coarse-grained over sufficiently
large volumes, and thus are expected to be approximately decoherent
(because exactly conserved quantities are exactly decoherent in the histories
approach). This is therefore a different mechanism for decoherence than
the usual mechanism of decoherence through an environment.
Hence, in order to close
the gap between the familiar system--environment models and the
less familiar hydrodynamic models without an obvious environment,
it is useful to rewrite the system--environment models in terms of
local densities as we do here.

In Section 2, we briefly review many body field theory, and
carry out the simple scattering calculation leading to the result
the the scattering particles effectively measure the local
number density, Eq.(\ref{1.12}).

In Section 3, we use many body field theory to derive the master
equation for the system, using a slow motion limit for the gas of
massive particles. As anticipated it has the Lindblad form in with
the Lindblad operators proportional to the Fourier-transformed
number density $N_{\k} $.

In Section 4 we show that our master equation reduces to an earlier
result of Gallis and Fleming \cite{GaF} in the one-particle sector for
the gas of massive particles. (This is essentially the same as the
master equation of Joos and Zeh \cite{JoZ} but the comparison with
Gallis and Fleming is more direct).

The master equation of Sections 3 and 4 does not have any dissipation
and is analogous to the Lindblad equation of quantum Brownian motion
with $L$ proportional to $x$.
In Section 5, we go beyond the slow-motion limit to derive a master equation
with dissipative terms.

We summarize and conclude in Section 6.

\section{Many Body Field Theory}

The dynamics of a many body system is very conveniently handled
using many body quantum field theory.
We now set up the formalism of many body field theory \cite{Zub,FeWa}
which we will use
to derive the master equation.
We consider a set of non-relativistic system particles described by a field
$\psi (\x)$ interacting through a potential $\phi (\x)$ with an environment
described by a field $\chi (\x) $. The total system is described by the Hamiltonian
\begin{eqnarray}
H &=& \int d^3 x \ \left( { 1  \over 2M} \nabla \psi^{\dag} (\x) \cdot \nabla \psi (\x)
+  { 1 \over 2m} \nabla \chi^{\dag} (\x) \cdot \nabla \chi (\x) \right)
\nonumber \\
&+& \half \int d^3x d^3 x' \ \psi^{\dag} (\x ) \psi (\x') \phi (\x
- \x') \chi^{\dag} (\x') \chi (\x)
\label{2.1}
\end{eqnarray}
(For simplicity we set $\hbar = 1$ hereafter).
In this language, the number densities $N(\x )$ and $n(\x )$ of the system and environment
fields are
\begin{eqnarray}
N (\x ) &=& \psi^{\dag} (\x) \psi (\x)
\\
n (\x ) &=& \chi^{\dag} (\x) \chi (\x)
\label{2.2}
\end{eqnarray}
(This is the field-theoretic version of Eq.(\ref{1.12})).

The above relations are also more conveniently written in terms of
$a_{\k}$ and $b_{\k}$, the annihilation operators for
the system and environment, respectively, and the Hamiltonian then is
\begin{eqnarray}
H &=& \sum_\q \left( E_{\q} a^{\dag}_{\q} a_{\q}
+ \omega_{\q} b^{\dag}_{\q} b_{\q} \right)
\\\nonumber
&+& {1 \over 2 V} \sum_{\k_1' + \k_2' = \k_1 + \k_2}
\nu ( \k_2' - \k_2 ) a^{\dag}_{\k_1} b^{\dag}_{\k_2}
a_{\k_1'} b_{\k_2'}
\label{2.3}
\end{eqnarray}
where $ E_{\q} = \q^2 / 2 M $, $\omega_{\q} = \q^2 / 2 m $,
$V$ is the spatial volume of the system (which we assume is in a box) and
\begin{equation}
\nu (\k) = \int d^3x \ e^{ - i \k \cdot \x} \ \phi (\x)
\end{equation}
The Fourier
transformed number densities are
\begin{eqnarray}
N_\k  &=& \sum_{\q} a^{\dag}_{\q} a_{\q + \k}
\\
n_\k  &=& \sum_{\q} b^{\dag}_{\q} b_{\q + \k}
\label{2.4}
\end{eqnarray}
and one may see that the Hamiltonian has the more concise form
\begin{eqnarray}
H & = &\sum_\q \left( E_{\q} a^{\dag}_{\q} a_{\q}
+  \omega_{\q} b^{\dag}_{\q} b_{\q} \right)
+ {1 \over 2 V} \sum_{\k} \nu (\k) N_{\k} n_{-\k}
\\
&=& H_0 + H_{int}
\label{2.5}
\end{eqnarray}
From these relations we see that the environment couples to the number density of the
system. It is this feature of many body field theory that makes it the appropriate
medium for the derivation of the master equation emphasizing the role of number density.

The $S$-matrix is
\begin{equation}
S = T \exp \left( - i \int_{ - \infty}^{\infty} dt \ H_{int} (t) \right)
\label{2.6}
\end{equation}
where
\begin{equation}
H_{int} (t) = {1 \over 2 V} \sum_{\k} \nu (\k) N_{\k} (t)  n_{-\k} (t)
\label{2.7}
\end{equation}
and here
\begin{eqnarray}
N_\k (t) &=& \sum_{\q} a^{\dag}_{\q} a_{\q + \k} \ e^{ i (E_\q - E_{ \q + \k})t}
\\
n_\k  (t)  &=& \sum_{\q} b^{\dag}_{\q} b_{\q + \k} \ e^{ i (\omega_{\q} - \omega_{\q + \k} ) t }
\label{2.8}
\end{eqnarray}

We may now use this formalism to look at a simple scattering situation to determine
how the environment stores information about the system. In the quantum Brownian motion
case, the nature of information storage was determined in essence by solving the
classical equations of motion. A similar strategy works here too. Let us suppose
the distinguished system is classical, and consider what happens when the environment
scatters off it. Suppose the environment starts in an initial momentum
state $ | \k_0 \rangle $ and scatters into a final state $ | \k_f \rangle$.
The scattering amplitude for this process, to first order, is
\begin{eqnarray}
\langle \k_f | S | \k_0 \rangle  &=& { i \over 2 V}  \int_{ - \infty}^{\infty}
dt \sum_{\k} \nu (\k) N_{\k} (t)
\langle \k_f | n_{-\k} (t) | \k_0 \rangle
\nonumber \\
&=& {i \over 2 V} \ \nu (\k ) \ \int dt \  \ N_{\k} (t) \ e^{ i (\omega_{\k_f} -
\omega_{\k_0} ) t }
\label{2.9}
\end{eqnarray}
where $ \k = \k_f - \k_0 $. This simple result shows that a single scattering event by the
environment stores information about the Fourier transform (in space and time) of the number
density. It is in this sense that the number density has a preferred status -- this is
the variable that is measured most directly by the environment and is the exact analogy
of the relations Eqs.(\ref{1.11a}), (\ref{1.11b}) in the quantum Brownian case.

The measured variables above are of course non-local in time, involving a temporal Fourier
transform of the number density, so cannot in fact be compatible with a Markovian
master equation.
Under a reasonable slow motion assumption, the system timescale
is much slower than the environment
timescale, and we may ignore the time-dependence in $ N_{\k} (t)$, yielding
\begin{equation}
\langle \k_f | S | \k_0 \rangle
= {i \over 2 V} \ \nu (\k ) \   \ N_{\k} \ \delta (\omega_{\k_f} - \omega_{\k_0} )
\label{2.10}
\end{equation}
This corresponds more directly to a Markovian master equation, as we shall see.

It remains to briefly sketch the connection between these results
and the discussion in Section 1 of records in the decoherent
histories approach. We imagine that the environment consists of a
very large number of particles which scatter off the system
particles. Each scattering event consists of an incoming
environment particle with momentum $\k_0$ scattered into a final
state of momentum $\k_f$, as outlined above. After the scattering
event, which is essentially instantaneous (on the timescale of
system dynamics), we may imagine that the scattered particle
propagates freely and may be measured at any time in the future.
Therefore, the records in the decoherent histories approach
consist of projections at the end of this histories onto the
momenta of all the scattered environment particles, from which the
number densities $N_{\k}$ of the system at a series of times may
be retrodicted.

\section{Derivation of the Master Equation}

Following the method first used by Joos and Zeh \cite{JoZ}, we may derive
the master equation for the reduced density operator $\rho$ of the system
by considering the scattering of the environment off the system, to second
order in interactions. We assume that the system and environment are initally
uncorrelated, so the total density operator is
\begin{equation}
\rho_T = \rho_0 \otimes \rho_{\E}
\label{3.1}
\end{equation}
We also assume that each scattering event takes place on a timescale
which is extremely short compared to the timescale of system dynamics. This means
that in an interval of time $\Delta t $ which is long for the environment
but short for the system, we may write,
\begin{equation}
\rho_T (t + \Delta t ) = S \rho_T (t) S^{\dag}
\label{3.2}
\end{equation}
(where we are using the interaction picture).
Expanding (\ref{3.2}) to second order, the $S$-matrix may be
written,
\begin{equation}
S = 1 + i U_1 - U_2
\label{3.3}
\end{equation}
where
\begin{equation}
U_1 =   - \int_{ - \infty}^{\infty} dt \ H_{int} (t)
\label{3.4}
\end{equation}
and
\begin{equation}
U_2 = \half \int dt_1 \int dt_2 \ {\rm T} \left( H_{int} (t_1) H_{int} (t_2) \right)
\label{3.5}
\end{equation}
The requirement of unitarity, $ S^{-1} = S^{\dag}$, implies that $ U_1 = U_1^{\dag}$
and
\begin{equation}
U_2 + U_2^{\dag} = U_1^2
\label{3.6}
\end{equation}
We will therefore write
\begin{equation}
U_2 = \half U_1^2 + i B
\label{3.7}
\end{equation}
where $B = B^{\dag}$, so we now have
\begin{equation}
S = 1 + i (U_1 - B) - \half U_1^2
\label{3.8}
\end{equation}
Inserting this in (\ref{3.2}), we obtain
\begin{equation}
{ d \rho_T \over dt} \ \Delta t = i [ U_1 - B , \rho_T ]
+ U_1 \rho_T U_1 - \half U_1^2 \rho_T - \half \rho_T U_1^2
\label{3.9}
\end{equation}
We now trace Eq.(\ref{3.9}) over the environment to obtain the master
equation for the system density operator $\rho$. As is usual in this
sort of model, we assume that the environment is so large that
its state is essentially unaffected by the interaction with the system.
Since the total density operator starts out in the factored
state (\ref{3.1}), this then means that, to a good approximation,
$\rho_T$ persists in the approximately factored form
$ \rho \otimes \rho_{\E}$, and we may insert this
in the right-hand side of Eq.(\ref{3.9}) \cite{Gar}.
We thus obtain the preliminary form for the master equation
\begin{equation}
{ d \rho \over dt} \ \Delta t = i [ {\rm Tr}_\E (U_1 \rho_{\E}) - {\rm Tr}_\E ( B \rho_\E), \rho ]
+ {\rm Tr}_{\E} \left(  U_1 \rho_T U_1 - \half U_1^2 \rho_T - \half \rho_T U_1^2 \right)
\label{3.10}
\end{equation}

We now work out these terms in more detail. We first consider the simple but
useful slow motion approximation, in which we ignore the time-dependence of $N_{\k} (t)$.
This implies that
\begin{equation}
U_1 \approx  - { 1 \over 2 V} \sum_{\k} \nu (\k) N_{\k}
\ \sum_{\q} b^{\dag}_{\q} b_{\q - \k} \  2 \pi   \delta (\omega_{\q} - \omega_{\q - \k} )
\label{3.11}
\end{equation}
The important terms for decoherence are the final three terms on the right-hand
side of (\ref{3.11}). When traced, these give,
\begin{equation}
{\rm Tr}_{\E} \left( U_1 \rho_T U_1 - \half U_1^2 \rho_T - \half \rho_T U_1^2 \right ) =
\sum_{\k \k'} c(\k, \k') \left( N_{\k'}  \rho N_{\k}
- \half N_{\k} N_{\k'} \rho - \half \rho N_{\k} N_{\k'} \right)
\label{3.12}
\end{equation}
where
\begin{equation}
c(\k, \k') = \nu (\k) \nu (\k')
\sum_{\q \q'}
 \ \delta (\omega_{\q} - \omega_{\q - \k} )  \delta (\omega_{\q'} - \omega_{\q' - \k' } )
\ \langle b^{\dag}_{\q} b_{\q - \k}  b^{\dag}_{\q'} b_{\q' - \k'} \rangle_{\E}
\label{3.13}
\end{equation}
We will take the environment to be a thermal state, which is diagonal in the
momentum states. It follows that
\begin{equation}
\langle b^{\dag}_{\q} b_{\q - \k}  b^{\dag}_{\q'} b_{\q' - \k'} \rangle_{\E}
\propto \delta_{\q, \q' - \k'} \ \delta_{\q', \q- \k}
\label{3.14}
\end{equation}
This implies $\k = - \k'$, and also that the two delta-functions are the same
in Eq.(\ref{3.13}).
We then interpret the square of the delta-function in the usual way,
\begin{eqnarray}
\left[ \delta (\omega_{\q} - \omega_{\q - \k} ) \right]^2 &=& \delta (0)
\ \delta (\omega_{\q} - \omega_{\q - \k} )
\nonumber \\
&=& \frac {\Delta t } {2 \pi}  \ \delta (\omega_{\q} - \omega_{\q - \k} )
\label{3.14}
\end{eqnarray}
We now have
\begin{equation}
c(\k, \k') = \delta_{\k, - \k'} \ c(\k) \frac{\Delta t} {2 \pi}
\label{3.14b}
\end{equation}
where
\begin{eqnarray}
c(\k ) &=&  \frac {1} {2 \pi} \left | \nu( \k) \right|^2 \sum_\q \   \ \delta (\omega_{\q} - \omega_{\q - \k} )
\langle b^{\dag}_{\q} b_{\q - \k}  b^{\dag}_{\q- \k} b_{\q} \rangle_{\E}
\nonumber \\
&=& \frac {1} {2 \pi} \left | \nu( \k) \right|^2 \sum_\q \   \ \delta (\omega_{\q} - \omega_{\q - \k} )
\langle b^{\dag}_{\q} b_{\q} \rangle_{\E} \left( \langle b^{\dag}_{\q - \k} b_{\q- \k } \rangle_{\E}
+ 1 \right)
\label{3.15}
\end{eqnarray}
The terms involving environment averages have the usual thermal form (for a bosonic environment),
\begin{equation}
\langle b^{\dag}_{\q} b_{\q} \rangle_{\E} = \frac {1} { e^{\beta (\omega_\q - \mu) } - 1 }
\end{equation}
where $\b = 1/ k T $ with $T$ temperature, and $\mu$ is the chemical potential.

The form Eq.(\ref{3.14b}) means that the important terms in the master equation are of the Lindblad form,
\begin{equation}
{ \rm Tr}_{\E} \left( U_1 \rho_T U_1 - \half U_1^2 \rho_T - \half \rho_T U_1^2 \right ) =
\Delta t \ \sum_{\k } c(\k ) \left( N_{\k}  \rho N^{\dag}_{\k}
- \half N^{\dag}_{\k} N_{\k} \rho - \half \rho N^{\dag}_{\k} N_{\k} \right)
\label{3.16}
\end{equation}
where we have used the fact that $ N^{\dag}_{\k} = N_{-\k} $.
The remaining two terms in Eq.(\ref{3.10}) clearly just modify the unitary
dynamics of the system. First we have
\begin{equation}
{\rm Tr}_\E \left( U_1 \rho_\E \right) = { 1 \over 2 V} \sum_{\k} \nu (\k) N_{\k}
\ \sum_{\q} \langle b^{\dag}_{\q} b_{\q - \k} \rangle_{\E} \  2 \pi   \delta (\omega_{\q}
- \omega_{\q - \k} )
\label{3.17}
\end{equation}
Clearly from the term $ \langle b^{\dag}_{\q} b_{\q - \k} \rangle_{\E} $ this
expression will be zero unless $\k = 0$, and therefore it is proportional to
$N$, the total particle number operator (although the overall coefficient will need
to be regularized). This therefore contributes a term to the master equation
of the form $ [ N, \rho ] $. We assume that there is a fixed number of system
particles so it is reasonable to take this term to be zero.

The other remaining term in Eq.(\ref{3.10}) involves the time ordering terms in $U_2$ and is a
bit more complicated
to evaluate. Fortunately, the detailed form of this expression is not needed here,
and it can in fact be easily shown that this term has the form
\begin{equation}
{\rm Tr}_\E ( B \rho_\E) = \Delta t \sum_{\k} \ d(\k) \ N_{\k} N^{\dag}_{\k}
\label{3.18}
\end{equation}
for some coefficient $d (\k) $ which we will not need.
Inserting all these
results in Eq.(\ref{3.10}), the factors of $\Delta t$ all drop out, and we
obtain, in the Schr\"odinger picture,
\begin{equation}
{ d \rho \over d t } = -i [ H_0 - \sum_{\k} \ d(\k)  \ N_{\k} N^{\dag}_{\k}, \rho]
+ \sum_{\k } c(\k ) \left( N_{\k}  \rho N^{\dag}_{\k}
- \half N^{\dag}_{\k} N_{\k} \rho - \half \rho N^{\dag}_{\k} N_{\k} \right)
\label{3.19}
\end{equation}
As desired, this is the Lindblad form with the Lindblad operators given
by
\begin{equation}
L_{\k} = c^{\half} (\k) N_{\k}
\end{equation}

We have therefore produced a derivation of the master equation for a scattering
environment which shows very clearly the connection between the preferred
basis (diagonalization in the Lindblad operators) and the information storage
about the system, as indicated by the simple scattering calculation, Eq.(\ref{2.10}).

It is interesting to note that the decoherence effect is second order in
interactions, but we were able to anticipate it from the simple first
order calculation, Eq.(\ref{2.10}). The reason for this is the relationship
Eq.(\ref{3.7}), which shows that the important part of the second order
terms is the square of the first order terms, and this is a consequence of unitarity.

\section{Comparison with Previous Works}

It is useful to check that the master equation we have derived reproduces
known results when we restrict to the one-particle sector for the system.
We will compare the results of this to the derivation of Gallis and Fleming
\cite{GaF} (which is essentially the same as Joos and Zeh \cite{JoZ} and Diosi
\cite{Dio}).

In the one-particle sector we may work with a density matrix $\rho (\k, \k')
= \langle \k | \rho | \k' \rangle$,
or equivalently $\rho (\x, \y) $ in the position representation.
We use the relations
\begin{eqnarray}
\left[ N_{\mathbf{q}},a_{\mathbf{k}}^{\dagger} \right]
& = & a_{\mathbf{k-q}}^{\dagger}
\\
\left[ N_{\mathbf{q}},a_{\mathbf{k}} \right]
& = & -a_{\mathbf{k+q}}
\end{eqnarray}
These relations imply that
\begin{eqnarray}
N_{\mathbf{q}}\rho(\mathbf{k},\mathbf{k'}) N_{\mathbf{-q}}
&=& \rho(\mathbf{k-q},\mathbf{k'-q})
\\
N_{\mathbf{-q}}N_{\mathbf{q}}\rho(\mathbf{k},\mathbf{k'})
&=&\rho(\mathbf{k},\mathbf{k'})
\\
\rho(\mathbf{k},\mathbf{k'})N_{\mathbf{-q}}N_{\mathbf{q}}
&=& \rho(\mathbf{k},\mathbf{k'})
\end{eqnarray}
In the position representation, this means
\begin{equation}
N_{\mathbf{k}}\rho(\mathbf{x},\mathbf{y}) N_{\mathbf{-k}}=
e^{i\mathbf{k}.(\mathbf{x}-\mathbf{y})}\rho(\mathbf{x},\mathbf{y})
\end{equation}
The master equation for the one-particle density operator $\rho (\x, \y )$ is
then
\begin{equation}
\frac { \partial \rho (\x, \y ) } {\partial t}
= - i \langle \x | \left[ H_0, \rho \right] | \y \rangle
- F (\x - \y ) \rho (\x, \y)
\end{equation}
where
\begin{equation}
F( \x - \y) = \frac{\pi} {2 V^2} \int d^3q d^3k \ \left| \nu
({\k}) \right|^2 \ n_{\mathbf{q}} (n_{\mathbf{q-k}}+1)\ \delta
(\omega_{\mathbf{q}}- \omega_{\mathbf{q-k}}) \ (1- e^{i\mathbf{k}
\cdot(\mathbf{x-y})}) \label{4.1}
\end{equation}
Note that the term involving the coefficient $d (\k) $ in Eq.(\ref{3.19}) drops
out because $ [ N_{\k} N^{\dag}_{\k} , \rho ] = 0$ in the one-particle sector.

To compare this with the Gallis-Fleming result \cite{GaF}, we first introduce
the quantity
\begin{equation}
f(\mathbf{k},\mathbf{k'})=\frac{ m } {2 \pi }
\nu ( \k - \k')
\end{equation}
(which appears in the usual Born approximation to first order scattering).
Then, letting $ \k \rightarrow -\k + \q $ in (\ref{4.1}), we get
\begin{equation}
F( \r) = \frac{ 4 \pi^2} {m^2}
\int d^3q d^3k
\ \left| f (\q, \k) \right|^2 \ n_{\mathbf{q}}
(n_{\mathbf{k}}+1)\ \delta (\omega_{\mathbf{q}}- \omega_{\mathbf{k}})
\ (1- e^{i(\mathbf{q-k})\cdot \mathbf{r}})
\label{4.2}
\end{equation}
The delta-function implies that $\q^2 = \k^2 $, and we find that
\begin{equation}
F( \r ) =  \frac{ 4 \pi^2} {m^2}
\int dq \ q^3 \ n_q ( n_q + 1 ) \ \int d\Omega d\Omega'
|f(\mathbf{q},\mathbf{k})|^2(1- e^{i (\mathbf{q-k}) \cdot \mathbf{r} })
\end{equation}
This in fact agrees with Gallis and Fleming if we identify
$  q / m $ as their $ v(q)$, the speed of the incoming particles,
and $ 4 \pi^3 q^2 n_q (n_q + 1 ) $ as the density of particles with speed
$q$.
In the one particle sector there is therefore agreement with earlier work.
(At least up to an overall numerical factor which we could not rectify. However,
we have also spotted some small and probably insignificant numerical errors in Ref.\cite{GaF}).

Mention should also be made of the master equations derived by
Unruh and Zurek, which used a field as an environment for a
particle \cite{UnZu}, and Anastopoulos and Zoupas \cite{AnZo},
which used a photon field and as an environment for a spinor
field. Also of relevance is the general account of the derivation
of master equations given by Omn\`es \cite{Omn1}. These works are
rather different to the present paper.

\section{Beyond the Slow Motion Approximation}

The derivation above assumed, in essence, that the system dynamics are infinitely
slow. Not surprisingly, the resulting master equation does not involve dissipation,
since, in the approximation used, the system is essentially at rest
for the timescale of a single
scattering event. It is analogous to the master equation of quantum Brownian
motion with the Lindblad operator $ L $ proportional to $x$, Eq.(\ref{1.14}).
To get a more realistic equation with dissipation
we therefore need to go beyond the slow motion approximation.

Because the local number density is a locally conserved quantity, it obeys
a continuity equation of the form,
\begin{equation}
\dot N_{\k} = - i \k \cdot \P_{\k}
\end{equation}
where $\P_\k$ is the local momentum density,
\begin{equation}
\P_{\k} =  \sum_{\q} \left( \q + \half \k \right)
a^{\dag} _{\q} a_{\q + \k}
\label{5.1}
\end{equation}
It is reasonable to expect that the master equation will involve this operator
when we go beyond the infinitely slow limit.
We now briefly repeat the derivation of the master equation, this time allowing
a slow time-dependence in $N_{\k} (t)$.

We have
\begin{equation}
U_1 = \frac {1} {2V} \sum_{\k} \nu (\k) \ \sum_{\q} b^{\dag}_{\q} b_{\q - \k}
\ \int dt \ N_{\k} (t) \ \ e^{ i (\omega_{\q} - \omega_{\q - \k} ) t }
\label{5.2}
\end{equation}
To take into account the time-dependence of $N_{\k} (t)$, we write,
\begin{equation}
N_{\k} (t) = N_{\k} + t \dot N_{\k} + \cdots
\end{equation}
where $\dot N_{\k}$ is given in terms of the momentum density, Eq.(\ref{5.1}).
Inserting this in Eq.(\ref{5.2}), the factor of $t$ may be rewritten in terms of a delta-function
derivative, yielding,
\begin{equation}
U_1 = \frac {1} {2V} \sum_{\k} \nu (\k) \ \sum_{\q} b^{\dag}_{\q} b_{\q - \k}
\left( N_\k \ \delta ( \omega_{\q} - \omega_{\q - \k} ) - i \dot N_{\k}
\ \delta '( \omega_{\q} - \omega_{\q - \k} ) + \cdots \right)
\end{equation}
We now use this expression for $U_1$ in the derivation of the master equation.
So for example, we get, in place of Eq.(\ref{3.12}),
\begin{equation}
{\rm Tr}_{\E} \left( U_1^2 \rho_{\E} \right)
= \Delta t \ \sum_{\k} \ c(\k ) \ N^{\dag}_{\k} N_{\k} \rho
+ i \sum_{\k} b(\k) \left( N_{\k} \dot N_{\k}^{\dag} - N_{\k}^{\dag} \dot N_{\k}
\right) \rho + \cdots
\end{equation}
where $c (\k) $ is given by Eq.(\ref{3.15}) and
\begin{equation}
b (\k) = \left | \nu( \k) \right|^2 \sum_\q \   \ \delta (\omega_{\q} - \omega_{\q - \k} )
\ \delta '(\omega_{\q} - \omega_{\q - \k} )
\ \langle b^{\dag}_{\q} b_{\q} \rangle_{\E} \left( \langle b^{\dag}_{\q - \k} b_{\q- \k } \rangle_{\E}
+ 1 \right)
\label{5.3}
\end{equation}
This coefficient may in fact be shown to be simply related to $c (\k) $.
The delta-function derivative may be dealt with by noting the formal
relation
\begin{equation}
 \delta (x) \delta '(x )
= \half \frac { \partial } { \partial x }
\left[ \delta (x  ) \right]^2
\label{5.4}
\end{equation}
Now note that
\begin{equation}
\omega_{\q} - \omega_{\q - \k} = \frac {1} {2m} \left( 2 \k \cdot \q - \k^2 \right)
\end{equation}
It follows that the delta function derivatives may be expressed in terms of
derivatives with respect to $q_i$ as
\begin{equation}
\delta ' (\omega_{\q} - \omega_{\q - \k} )
=  \frac {2 m } {\k^2} \ k_i \frac { \partial} { \partial q_i }
\delta  (\omega_{\q} - \omega_{\q - \k} )
\end{equation}
Inserting these relations in Eq.(\ref{5.3}) and integrating by parts yields,
\begin{equation}
b (\k) = - \half \left | \nu( \k) \right|^2 \sum_\q \   \ \left[ \delta (\omega_{\q} - \omega_{\q - \k} ) \right]^2
\frac {2 m } {\k^2} \ k_i \frac { \partial} { \partial q_i }
\ \left( \langle b^{\dag}_{\q} b_{\q} \rangle_{\E} \left( \langle b^{\dag}_{\q - \k} b_{\q- \k } \rangle_{\E}
+ 1 \right) \right)
\end{equation}
where we will interpret the square of the delta-function as in Eq.(\ref{3.14}).

Now for simplicity work in the high temperature limit, so
\begin{equation}
\langle b^{\dag}_{\q - \k } b_{\q - \k } \rangle_{\E} \ \ll \ 1
\end{equation}
and
\begin{equation}
\langle b^{\dag}_{\q} b_{\q} \rangle_{\E} \ \approx \ e^{\mu \b} e^{ - \b \omega_{\q} }
\end{equation}
It follows that
\begin{eqnarray}
k_i \frac { \partial} { \partial q_i }  \langle b^{\dag}_{\q} b_{\q} \rangle_{\E}
&=& - \b k_i \ \frac { \partial \omega_{\q} } { \partial q_i } \ \langle b^{\dag}_{\q} b_{\q} \rangle_{\E}
\nonumber
\\
&=& - \b \ \frac {\k \cdot \q} {m}  \ \langle b^{\dag}_{\q} b_{\q} \rangle_{\E}
\end{eqnarray}
and we arrive at the very simple result,
\begin{equation}
b( \k) = \frac{\beta}{2} \ \Delta t \ c(\k)
\end{equation}
It is not difficult to see that we then arrive at a master equation which is once again
of the Lindblad form, but this time with Lindblad operators of the form
\begin{eqnarray}
L_{\k} &=& c^{\half} (\k) \left( N_{\k} - i \frac{\b}{2}  \dot N_{\k} \right)
\nonumber
\\
&=& c^{\half} (\k) \left( N_{\k} -  \frac{\b}{2}  \ \k \cdot \P_\k \right)
\end{eqnarray}
(up to terms of order $\b^2$, which can be dropped in the
approximation we are using). This is clearly closely analogous to
the QBM result, Eq.(\ref{1.14}). (A closely analogous formula
appears in Diosi's paper \cite{Dio}).

\section{Summary and Discussion}

We have given a derivation of the master equation describing
a many-body system interacting with a reasonably general class
of environments. The form of the master equation emphasizes the
central role of the local number density, which is the system
variable measured most directly by the environment in a scattering
situation.

We did not in fact give a specific form for the
interaction between the system and environment since it was not necessary
to illustrate the general points we are making. Some specific forms for
this interaction are discussed elsewhere \cite{GaF,JoZ}.

The derivation reduces to familiar results of Gallis and Fleming \cite{GaF},
Diosi \cite{Dio}, and Joos and Zeh \cite{JoZ}, when we restrict to the one-particle sector
of the many-body field theory. The many-body derivation confers some
advantages of the usual derivations (which consider scattering theory
in quantum mechanics) in that they avoid essentially classical assumptions
about fluxes of scattering particles. Our derivation also has the possibility
of being extended to a low temperature regime (and to Bose-Einstein condensation,
for example) and to fermionic environments, although we do not discuss this here.

\section{Acknowledgements}

We are grateful to Lajos Diosi for useful conversations.
P.D. was supported by PPARC.

\bibliography{apssamp}

\begin{thebibliography}{10}




\bibitem{Zur1} See for example, W. Zurek, Physics Today {\bf 40}, 36 (1991)
% {\it Decoherence and the Transition from Quantum to Classical.}


\bibitem{GeH1} M.Gell-Mann and J.B.Hartle, Phys.Rev. {\bf D47},
3345 (1993).
%{\it Classical Equations for Quantum Systems}.

\bibitem{Hal1} J.J.Halliwell, Phys.Rev. {\bf D60}, 105031
(1999).
% Somewhere in the Universe: Where is the information stored when
% histories decohere

\bibitem{Har1} J.B.Hartle, in {\it Proceedings of the 1992 Les Houches Summer
School, Gravitation et Quantifications}, edited by B.Julia and
J.Zinn-Justin (Elsevier Science B.V., 1995).
%{\it Spacetime Quantum Mechanics and the Quantum Mechanics of Spacetime.}


\bibitem{GeH2} M.Gell-Mann and J.B.Hartle, in {\it Complexity, Entropy
and the Physics of Information, SFI Studies in the Sciences of
Complexity}, Vol. VIII, W. Zurek (ed.) (Addison Wesley, Reading,
1990); and in {\it Proceedings of the Third International
Symposium on the Foundations of Quantum Mechanics in the Light of
New Technology}, S. Kobayashi, H. Ezawa, Y. Murayama and S. Nomura
(eds.) (Physical Society of Japan, Tokyo, 1990).
%{\it Quantum Mechanics in the Light of Quantum Cosmology.}

\bibitem{Gri} R.B.Griffiths, J.Stat.Phys. {\bf 36}, 219 (1984);
{ Phys.Rev.Lett.} {\bf 70}, 2201 (1993).

\bibitem{Omn} R. Omn\`es, J.Stat.Phys. {\bf 53}, 893 (1988);
{\bf 53}, 933 (1988); {\bf 53}, 957 (1988); {\bf 57}, 357 (1989);
Ann.Phys. {\bf 201}, 354 (1990); Rev.Mod.Phys. {\bf 64}, 339
(1992).

\bibitem{Hal2} J.J.Halliwell, in {\it Fundamental Problems in Quantum
Theory},  edited by D.Greenberger and A.Zeilinger, Annals of the
New York Academy of Sciences, Vol 775, 726 (1994).

\bibitem{Zur2} W.H.Zurek, Phil.Trans.R.Soc.Lond. {\bf A356}, 1793
(1998).
%``Decoherence, Einselection, and the
%Existential Interpretation (The Rough Guide)'', preprint
%quant-ph/9805065.

\bibitem{Zur4} W.H.Zurek, Phys.Rev. {\bf D24}, 1516 (1981);
{\bf D26}, 1862 (1982).

\bibitem{JoZ} E.Joos and H.D.Zeh,  Z.Phys. {\bf B59}, 223 (1985).

\bibitem{GiP} N.Gisin and I.C. Percival, J.Phys.{\bf A25},
5677 (1992); {\bf A26}, 2233 (1993); {\bf A26}, 2245 (1993); Phys.
Lett. {\bf A167}, 315 (1992).

\bibitem{HaZ} J.Halliwell and A.Zoupas,  Phys.Rev.
{\bf D52}, 7294 (1995); {\bf D55}, 4697 (1997).

\bibitem{GeH3} M.Gell-Mann and J.B.Hartle,
in {\it Quantum Classical Correspondence: Proceedings of the 4th
Drexel Symposium on Quantum Nonintegrability}, edited by D.H.Feng
and B.L.Hu (International Press, 1997). (Also available as e-print
gr-qc/9509054 (1995)).
% Strong decoherence

\bibitem{CaL} See for example, A.Caldeira and A.Leggett,
 Physica {\bf 121A}, 587 (1983). The literature on quantum
Brownian motion models is considerable. A selection is,
G.S.Agarwal, Phys.Rev. {\bf A3}, 828 (1971); Phys.Rev. {\bf A4},
739 (1971); H.Dekker, Phys.Rev. {\bf A16}, 2116 (1977); Phys.Rep.
{\bf 80}, 1 (1991); G.W.Ford, M.Kac and P.Mazur, J.Math.Phys. {\bf
6}, 504 (1965); H.Grabert, P.Schramm, G-L. Ingold, Phys.Rep. {\bf
168}, 115 (1988); V.Hakim and V.Ambegaokar, Phys.Rev. {\bf A32},
423 (1985); J.Schwinger, J.Math.Phys. {\bf 2}, 407 (1961);
I.R.Senitzky, Phys.Rev. {\bf 119}, 670 (1960). Some derivations of
the master equation of quantum Brownian motion are, B.L.Hu, J.Paz
and Y. Zhang, Phys.Rev. {\bf D45}, 2843(1992); {\bf D47},
1576(1993); J.J.Halliwell and T.Yu, Phys.Rev. {\bf 53}, 2012
(1996).

\bibitem{Dio} L.Diosi, Europhys. Lett. {\bf 30}, 63 (1995).
% gr-qc/9403046.
% QUANTUM MASTER EQUATION OF A PARTICLE IN A GAS ENVIRONMENT
% EUROPHYS LETT 30 (2): 63-68 APR 10 1995

\bibitem{GaF} M.R.Gallis and G.N.Fleming, Phys. Rev. {\bf
A 42}, 38-48 (1990).
% Environmental and spontaneous localization

\bibitem{Lin} G.Lindblad, Comm.Math.Phys. {\bf 48}, 119 (1976).

\bibitem{Dio2} L.Di\'osi, Europhys.Lett. {\bf 22}, 1 (1993).

\bibitem{Amb} V.Ambegaokar, Ber.Bunsenges.Phys.Chem. {\bf 95},
400 (1991). (This author in turn cites a private communication
from P.Pechukas.)

\bibitem{DGHP} L.Di\'osi, N.Gisin, J.Halliwell and I.C.Percival,
Phys.Rev.Lett. {\bf 74}, 203 (1995).

\bibitem{Caves} C.M.Caves and G.J.Milburn, Phys.Rev. {\bf A36},
5543 (1987).

\bibitem{Hal3} J.J.Halliwell, Phys.Rev. {\bf D58}, 105015 (1998);
{ Phys.Rev.Lett. }{\bf 83}, 2481 (1999).

\bibitem{Zub} D.N.Zubarev, {\it Nonequilibrium Statistical
Thermodynamics} (Consultants Bureau, New York, 1974).

\bibitem{FeWa} A.L.Fetter and J.D.Walecka, {\it Quantum Theory
of Many-Particle Systems} (McGraw-Hill, 1971).

\bibitem{Gar} C.W.Gardiner, {\it Quantum Noise} (Springer-Verlag,
Berlin, 1991).


%\bibitem{Har6} J.B.Hartle, in, {\it Proceedings of
%the Cornelius Lanczos International Centenary Confererence},
%edited by J.D.Brown, M.T.Chu, D.C.Ellison and R.J.Plemmons
%(SIAM, Philadelphia, 1994)
%``Quasiclassical Domains in a Quantum
%Universe'', preprint gr-qc/9404017 (1994).

\bibitem{UnZu} W.G.Unruh and W.H.Zurek, Phys.Rev. {\bf D40},
1071 (1989).

\bibitem{AnZo} C.Anastopoulos and A.Zoupas, Phys.Rev.
{\bf D58}, 105006 (1998).
% Non-Equilibrium Quantum Electrodynamics Authors:

\bibitem{Omn1} R.Omn\`es, Phys.Rev. {\bf A56}, 3383 (1997).
% General theory of the decoherence effect in quantum mechanics.

%\bibitem{Zur0} See for example, J.P.Paz and W.H.Zurek, { Phys.Rev.} {\bf D48},
%2728 (1993); W.Zurek, in {\it Physical Origins of Time Asymmetry},
%edited by  J.J.Halliwell, J.Perez-Mercader and W.Zurek (Cambridge
%University Press, Cambridge, 1994). Zurek's contributions
%to the subject are extensive and the above is only a
%suggestive selection.

%\bibitem{Zur3} An interesting collection of article on many aspects of
%the physics of information may be found in, {\it Complexity, Entropy
%and the Physics of  Information}, edited by W.Zurek
%(Addison--Wesley, Redwood City, CA, 1990).


\end{thebibliography}

\end{document}